# Systematic calculation on α decay and cluster radioactivity of superheavy nuclei

Xuanpeng Xiao[1], Panpan Qi[1], Gongming Yu[1], Haitao Yang[2], Qiang Hu[3]

[1]College of Physics and Technology, Kunming University, Kunming 650214, China
[2]College of Science, Zhaotong University, Zhaotong 657000, China
[3]Institute of Modern Physics, Chinese Academy of Sciences, Lanzhou 730000, China

In the Coulomb and Proximity Potential Model (CPPM) framework, we have investigated the cluster radioactivity and α decay half-lives of superheavy nuclei. We study 22 different versions of proximity potential forms that have been proposed to describe proton radioactivity, two-proton radioactivity, heavy-ion radioactivity, quasi-elastic scattering, fusion reactions, and other applications. The half-lives of cluster radioactivity and α decay of 41 atomic nuclei ranging from 221Fr to 244Cm were calculated, and the results indicate that the refined nuclear potential named BW91 is the most suitable proximity potential form for the cluster radioactivity and α decay of superheavy nuclei since the root-mean-square (RMS) deviation between the experimental data and the relevant theoretical calculation results is the smallest ($\sigma$ = 0.841). By using CPPM, we predicted the half-lives of 20 potential cluster radioactivity and α decay candidates. These cluster radioactivities and α decays are energetically allowed or observable but not yet quantified in NUBASE2020.

Keywords: α decay, cluster radioactivity, half-lives, Coulomb and proximity potential model

PACS number(s): 23.60.+e, 23.70.+j, 27.80.+w, 27.90.+b

## I. INTRODUCTION

The decay of superheavy nuclei(SHN) has always been an important field in nuclear physics. The primary decay modes of SHN include α decay, cluster radioactivity, and spontaneous fission[1]. In recent years, α decay still occupies an important place in nuclear physics. As one of the main decay modes of SHN, α decay is considered a simple decay process compared to other decay modes[2-4]. The process of α decay is essentially a quantum tunnelling effect, which was first explained by Gamow and Condon and Guerney in the 1920s[5-6]. Cluster radioactivity is the spontaneous emission of particles heavier than α particles and can be treated as an intermediate process between α decay and spontaneous fission[7]. This decay mode has garnered significant interest among physicists due to its ability to provide substantial and critical information for the study of nuclear structure. Moreover this phenomenon was first predicted by Sandulescu et al in 1980[8] and was experimentally observed by Rose and Jones[9]. The α decay of atomic nuclei can be investigated by using a variety of theoretical approaches, such as the generalized liquid dropmodel(GLDM)[10,11], the effective liquid drop model(ELDM)[12], the modified generalized liquid drop model(MGLDM)[13-15], Coulomb proximity potential model(CPPM)[16], and the preformed cluster model(PCM)[17]. These models use diverse interaction potentials such as the Proximity Potential[18], the Woods-Saxon Potential, the Squared Woods-Saxon Potential, and the Cosh Potential, as well as the bifold Potential at the microscopic level[19].

In recent years, Ren[20] has calculated the nuclear potential through a bifold modeling approach under the theoretical framework of the microscopic density-dependent model(DDCM). In this process, they used the effective M3Y nucleon-nucleon interaction to describe the nuclear density and calculated the cluster radioactivity halflife. Recently, Li[21] has calculated the α decay and cluster radioactivity SHN in the well-known 28 different versions of proximity potential formalisms and predicted multiple cluster radioactivity candidates. By introducing the branching ratio of cluster radioactivity relative to α decay, it can be found that the predicted atomic nuclei are more likely to undergo α decay than cluster radioactivity[21]. Moreover, some empirical or semiempirical formulas can be used to calculate the half-life of cluster radioactivity, such as the Akrawy formula[22], the universal decay law(UDL) proposed by Qi[23], Brown's scaling law, and the cluster radioactivity scaling law(SL) proposed by Horoi[24].

In the present work, the nuclear density distribution is regarded as a spherical distribution which neglects any effect of the charge distribution inside a finite-size neutron, then the nuclear charge density can be obtained by folding the product of the point proton density and the single-proton charge distribution[25]. Considering the formation probability with a simple mass dependence on the emission cluster, we systematically investigated the half-lives of cluster radioactivity and α decay for 41 nuclei using CPPM with 22 different versions of proximity potential formalisms. The calculation results indicate that BW91 has the smallest RMS deviation in the study of cluster radioactivity.

This paper is organized as follows. The theoretical framework of CPPM with 22 different proximity potential formalisms and the Ni's empirical formulae are presented in Section II.The results and discussion are presented in Section III. Finally, Section IV is a summary section.

## II. THEORETICAL FRAMEWORK

### A. Half-lives of the cluster radioactivity

The Coulomb potential $V_C(r)$ was calculated using the bifold model[26-30], which is expressed as:

$$V_c(\vec{r}) = \int\int \rho_c(\vec{r}_c)\rho_d(\vec{r}_d)v(s)d\vec{r}_c d\vec{r}_d, \quad (1)$$

where $\rho_c(\vec{r}_c)$ and $\rho_d(\vec{r}_d)$ are the matter and charge density distributions of the cluster and daughter nucleus, respectively. v(s) represents the proton-proton interaction. By neglecting the effect from the charge distribution inside a finite-size neutron, the nuclear charge density can be derived by folding the point proton density with the single-proton charge distribution[25,31]:

$$\rho(r) = \int \rho_p(\vec{r}\,')\rho^p(|\vec{r}-\vec{r}\,'|)d\vec{r}\,', \quad (2)$$

where $\rho^p$ takes the Gaussian form as follows[32-34]:

$$\rho^p = \frac{1}{(\sqrt{\pi} a_p)^3} e^{\frac{-r^2}{a_p^2}}, \quad (3)$$

where $a_p = 0.65$ fm[35], v(s) represents the proton-proton potential:

$$V(S) = \frac{e^2}{4\pi\varepsilon_0}\frac{1}{s}, \quad (4)$$

where $e^2 = 1.44$MeV·fm, and the proton is assumed to be a point-like particle, and $s = |\vec{r}_c - \vec{r}_d + \vec{r}|$ is the relative distance between interacting nucleon pair[36].

In the CPPM, the total inter action potential V(r) between the parent and the daughter nuclei, which includes the nuclear potential $V_N(r)$, the Coulomb potential $V_c(r)$ and the centrifugal potential $V_l(r)$, can be expressed as[37]:

$$V_l(r) = V_N(r) + V_c(r) + V_l(r), \quad (5)$$

The Langer modified centrifugal potential is also defined as[38]:

$$V_l(r) = \frac{(l+1/2)^2\hbar^2}{2\mu r^2}, \quad (6)$$

where $\hbar$ is the reduced Planck constant, and the reducedmass is $\mu = m\frac{A_c A_d}{A}$, $m = 931.5\mathrm{MeV/c^2}$, $A_c$ and $A_d$ denote the mass numbers of the emitted cluster and daughter nuclei, respectively, andAis mass number of the parent nucleus[39]. We adopt the Langer correction form $l(l+1) \to (l+1/2)^2$ to ensure the validity of the WKB approximation. Since the radius approaches zero after the Langer modified, the leading behavior exactly matches that of the exact solution. Otherwise, the WKB solution only matches the leading term of the exactsolution through the first two orders in $\hbar$[40,41]. The angular momentum $l$ carried by the emitted cluster can be obtained by standard selection rules[42,43]:

$$l = \begin{cases} \Delta_j, & \text{for even } \Delta_j \text{ and } \pi_p = \pi_d, \\ \Delta_{j+1}, & \text{for even } \Delta_j \text{ and } \pi_p \neq \pi_d, \\ \Delta_j, & \text{for odd } \Delta_j \text{ and } \pi_p \neq \pi_d, \\ \Delta_{j+1}, & \text{for odd } \Delta_j \text{ and } \pi_p = \pi_d, \end{cases} \quad (7)$$

where $\pi_p$ and $\pi_d$ are the parity value of the parent nucleus and the daughter nucleus, respectively. $\Delta_j = |j_p - j_d - j_c|$, and $j_p$, $j_d$, $j_c$ are the isospin value of the parent nuclei, the daughter nuclei and the emitted cluster, respectively[44].

The partial half-life is related to the decay constant λ by[45]

$$T_{1/2}=\frac{\ln 2}{\lambda}, \quad (8)$$

In the microscopic approach, λ can be expressed as[46]:

$$\lambda = vS_cP, \quad (9)$$

here, $v = 1.0 \times 10^{22}s^{-1}$ is the frequency of the assault on the barrier[47-49]. $S_c$ is the preformation probability of an α particle on the surface of the nucleus[50], which is given by overlaps of the nucleon states in the cluster with those of the parent nucleus[51] and can be expressed as:

$$S_c = S_i^{(A_c-1)/3}, \quad (10)$$

where $S_i(i = even, odd)$ is the spectroscopic factor for the α decay[49,51-53]. In this study, we choose $S_{even} = 0.02897$ for even-even parent nuclei and $S_{odd} = 0.0214$ for odd-A parent nuclei[54].

The barrier penetrability corresponding to the separated fragments is calculated in the semiclassical Wentzel-Kramers-Brillouin(WKB) approximation

$$P = \exp(-\frac{2}{\hbar}\int_{R_{in}}^{R_{out}} \sqrt{2\mu|V_r - Q_c|dr}), \quad (11)$$

Where $Q_c$ represents released energy[55],

$$Q_c = B(A_c, Z_c) + B(A_d, Z_d) - B(A, Z), \quad (12)$$

where $B(A_c, Z_c)$, $B(A_d, Z_d)$ and $B(A, Z)$ are the binding energies of the emitted cluster, daughter nucleus, and parent nucleus, respectively[21]. $Z_c$, $Z_c$ and Z are the proton numbers of the emitted cluster, daughter nucleus, and parent nucleus, respectively. The classical turning points $R_{in} = 1.2249(A_c^{1/3} + A_d^{1/3})$ and

$R_{out} = (\sqrt{(\frac{Z_cZ_de^2}{2Q_c})^2 + \frac{\hbar^2(l+1/2)^2}{2\mu Q_c}} + (\frac{Z_cZ_de^2}{2Q_c})^2)$[47,52].

## B. Proximity potential formalism

### 1. Proximity potential 77 family

The original version of the proximity potential for two spherical interacting nuclei could be expressed as[57,58]

$$V_N(r) = 4\pi\gamma b\bar{R}\Phi(\xi), \quad (13)$$

where $(\gamma b\overline{R})$ refers to the geometry of the participant nuclei[58], and $\mathrm{b} = 1\mathrm{fm}$. Surface energy coefficient $\gamma$ has the following form[59]:

$$\gamma = \gamma_0(1 - k_s A_s^2), \quad (14)$$

where $A_s = \frac{N_p - Z_p}{N_p + Z_p}$ is the asymmetry parameter and refers to neutron-proton excess[57], where $N_p$ and $Z_p$ are the neutron and proton numbers of the parent nucleus, respectively. $\gamma_0$ and $K_s$ are the surface energy constant and the surface asymmetry constant, respectively. $\gamma_0 = 1.01734\mathrm{MeV/fm}^2$ and $\mathrm{k_s} = 1.79$ [60]. Their details are listed in Table 1. $\overline{R}$ is the mean curvature radius or reduced radius

$$\overline{R} = \frac{C_c C_d}{C_c + C_d}, \quad (15)$$

where

$$C_i = R_i[1 - (\frac{b}{R_i})^2] \quad (i = c, d), \quad (16)$$

The effective sharp radius $R_i$ is defined as

$$R_i = 1.28A_i^{1/3} - 0.76 + 0.8A_i^{-1/3} \, (i = c, d), \quad (17)$$

The parametrization of the universal function is as follows:

$$\Phi(\xi) = \begin{cases} -\frac{1}{2}(\xi - 2.54)^2 - 0.0853(\xi - 2.54)^3, \xi < 1.2511, \\ -3.437exp(-\frac{\xi}{0.75}), \quad \xi \geq 1.2511, \end{cases} \quad (18)$$

where $\xi = \frac{r - C_c - C_d}{b}$ is the distance between the near surface of the emitted cluster and daughter nucleus.

According to the different modifications of the adjustable parameters of Prox.77, we divide the related versions of Prox.77 into eight categories. These classifications are based on the adjustment or change of surface energy coefficient, kernel radius and general function. A total of 22 versions of the proximity formalism are included in this study: (i)Prox.77 family and its modified versions based on adjustment of the surface energy coefficient[59,67], (ii)Prox.1981(Prox.81)[68], (iii)Bass family (including Bass73[69] and its modified forms Bass77[70] and Bass80[71]), (iv)Winther family (including CW76[72] and its modified forms BW91[71] and AW95[73]), (v)Ng ô 80[74], (vi)Guo2013[75].

Table 1.Different sets of surface energy coefficients γ correspond to different surface energy constants $\gamma_0$ and surface asymmetry constants $k_s$.

| γset | $\gamma_0(MeV/fm^2)$ | $k_s$ | references |
|---|---|---|---|
| Set1(γ-MS1967) | 0.9517 | 1.7826 | [59] |
| Set2(γ-MS1966) | 1.01734 | 1.79 | [60] |
| Set3(γ-MS1976) | 1.460734 | 4.0 | [61] |
| Set4(γ-KN S1969) | 1.2402 | 3.0 | [62] |
| Set5(γ-MN-I1981) | 1.1756 | 2.2 | [63] |
| Set6(γ-MN-II1981) | 1.27326 | 2.5 | [63] |
| Set7(γ-MN-III1981) | 1.2502 | 2.4 | [63] |
| Set8(γ-RR1984) | 0.9517 | 2.6 | [64] |
| Set9(γ-MN 1988) | 1.2496 | 2.3 | [65] |
| Set10(γ-MN 1995) | 1.25284 | 2.345 | [66] |
| Set11(γ-PD-LDM2003) | 1.08948 | 1.9830 | [67] |
| Set12(γ-PD-NLD2003) | 0.9180 | 0.7546 | [67] |
| Set13(γ-PD-LSD2003) | 0.911445 | 2.2938 | [67] |

2. Proximity potential Prox.81

Proximity potential Prox.81 was proposed by Blocki and Swiatecki in 1981[68]. It has the same form as Prox.77, except for the surface energy coefficient $\gamma = 0.9517[1 - 1.7826(\frac{N_p - Z_p}{A_p})^2]$ and universal function $\Phi(\xi = \frac{r - C_c - C_d}{b})$ written as

$$\Phi(\xi) = \begin{cases} -1.7817 + 0.9270\xi + 0.143\xi^2 - 0.09\xi^3, & \xi < 0, \\ -1.7817 + 0.9270\xi + 0.0.01696\xi^2 - 0.05148\xi^3, & \xi < 0, \\ -4.41exp(-\frac{\xi}{0.7176}), & \xi \geq 1.2511, \end{cases} \quad (19)$$

3. Proximity potential Bass family

In 1973, Bass, Based on the liquid drop model, obtained the proximity potential with the difference in surface energies between finite and infinite separation ξ[69]. Then, Bass proposed another form of proximity potential, marked as Bass77[70], and later proposed Bass80[71] proximity potential based on Bass77.Their details are listed in Table 2.

In Table 2, $\xi = r - R_c - R_d$. R is the sum of the half-maximum density radii and $r_0 = 1.07$ fm, as $a_s = 17$ MeV, $d = 1.35$ fm, $R = R_C + R_d$ in Bass73. $R_i = R_{si}(1 - \frac{0.98}{R_{si}^2})$

Table 2. Various versions of Bass family proximity potential $V_N$ and universal function $\Phi(\xi)$.

| | proximity potential | radius of the nucleus | universal function |
|---|---|---|---|
| Bass73 | $V_N = \frac{-da_s(A_c)^{1/3}(A_d)^{1/3}}{R}\exp(-\frac{r-R}{d})$ | $R_i = r_0(A_i)^{1/3}, (i=c,d)$ | \ |
| Bass77 | $V_N(r) = \frac{-R_cR_d}{R_c+R_d}\Phi(\xi)$ | $R_i = 1.16(A_i)^{1/3} - 1.39(A_i)^{1/3}, (i=c,d)$ | $\Phi(\xi) = [0.03\exp(\frac{\xi}{3.3}) + 0.0061\exp(\frac{\xi}{0.65})]^{-1}$ |
| Bass80 | $V_N(r) = \frac{-R_cR_d}{R_c+R_d}\Phi(\xi)$ | $R_{si} = 1.28(A_i)^{1/3} - 0.76 + 1.8(A_i)^{1/3}, (i=c,d)$ | $\Phi(\xi) = [0.033\exp(\frac{\xi}{3.3}) + 0.007\exp(\frac{\xi}{0.65})]^{-1}$ |

4. Proximity potential Winther family

In 1976, Christensen and Winter proposed an empirical nuclear potential called CW76 based on heavy ion elastic scattering data[71]. In 1991, Broglia and Winther proposed a nuclear potential named BW91[71] by using the Woods-Saxon parameterization of the proximity potential CW76. The nuclear potential AW95[73] is obtained by adjusting the parameters on the basis of BW91. Their details are listed in Table 3.

In Table 3, the parameter $\xi = r - R_c - R_d$, where R is the sum of the half-maximum density radii. $V_0 = 16\pi\gamma a\frac{R_cR_d}{R_c+R_d}$, $a = 0.63\text{fm}$, The expression and other parameters of the proximity potential of AW95 are the same as those of BW91, except for $a = \frac{1}{1.17(1+0.53(A_c^{-1/3}+A_d^{-1/3}))}$. In BW91, $R = R_c + R_d + 0.29$. In AW95 $R = R_c + R_d$.

Table 3. Various versions of Winther family proximity potential $V_N$ and universal function $\Phi(\xi)$.

| | proximity potential | radius of the nucleus | universal function |
|---|---|---|---|
| CW76 | $V_N(r) = -50\frac{R_cR_d}{R_c+R_d}\Phi(\xi)$ | $R_i = 1.233(A_i)^{1/3} - 0.978(A_i)^{1/3}, (i=c,d)$ | $\Phi(\xi) = \exp(\frac{-\xi}{0.63})$ |
| BW91 | $V_N(r) = \frac{-V_0}{1+\exp(\frac{r-R}{0.63})}$ | $R_i = 1.233(A_i)^{1/3} - 0.98(A_i)^{1/3}, (i=c,d)$ | \ |
| AW95 | $V_N(r) = \frac{-V_0}{1+\exp(\frac{r-R}{0.63})}$ | $R_i = 1.2(A_i)^{1/3} - 0.09(A_i)^{1/3}, (i=c,d)$ | \ |

5. Proximity potential Ngô80

In 1980, H. Ngô and Ch. Ngô derived a general formula for the nuclear component of the interaction potential between two heavy ions based on the proximity theorem.

$$V_N(r)=\frac{C_cC_d}{C_c+C_d}\Phi(\xi), \quad (20)$$

where $C_i=R_i[1-(\frac{b}{R_i})^2],(i=c,d)$ represents the central radii of the emitted cluster and daughter nucleus, where $R_i=\frac{N_iR_{ni}+Z_iR_{pi}}{A_i},(i=c,d)$ represents the sharp radii, and $R_{ij}=r_{0ij}A_i^{1/3},(j=p,n \quad i=c,d)$, where $r_{0pi}=1.128$fm and $r_{0ni}=1.1375+1.875\times10^{-4}A_i$fm. The universal function $\Phi(\xi=r-C_c-C_d)$ is given by

$$\Phi(\xi)=\begin{cases}-33+5.4(\xi+1.6)^2, & \xi<-1.6,\\ -33\exp(-\frac{\xi}{0.75}), & \xi\geq-1.6,\end{cases} \quad (21)$$

6. Proximity potential Guo2013

In 2013, Guo calculated the nuclear potential VN (Guo2013) using the bifold model with a density-dependent NN interaction[75]. It is expressed as:

$$V_N(r)=4\pi\gamma\frac{C_cC_d}{C_c+C_d}\Phi(\xi), \quad (22)$$

where $\gamma=0.9517[1-1.7826(\frac{N-p-Z_p}{A_p})^2]$ is the surface coefficient, and the universal function $\Phi(\xi)=\frac{P_1}{1+\exp(\frac{\xi+P_2}{P_3})}$ where $\xi=\frac{r-R_c-R_d}{b}$. $P_1=-17.72$, $P_2=1.30$, $P_3=0.854$are the adjustable parameters.$R_i=1.28A_i^{1/3}-0.8A_i^{1/3},(i=c,d)$is the effective sharp radius.

C. empirical formulas

1. Ni's empirical formula

In 2008, Ni derived some approximations from the WKB barrier penetration probability[76]and proposed a unified formula of half-lives for α decay and cluster radioactivity

$$\log_{10}^{T_{1/2}}=aZ_cZ_d\sqrt{U}Q_c^{1/2}+b\sqrt{U}(Z_cZ_d)^{1/2}+c, \quad (23)$$

where $Q_c$ represents the cluster radioactivity and α decay released energy. $U=A_cA_d/(A_c+A_d)$ is the reduced mass of the emitted cluster daughter nucleus system. For α decay, the adjustable parameters are $a=0.39961$, $b=-1.31008$, and $c=-$ $17.00698$ for even-even nuclei; $c=-16.26029$ for even-odd nuclei; and $c=-16.40484$ for odd-even nuclei; For cluster radioactivity, the adjustable parameters are $a=0.38617$, $b=-1.08676$, and $c=-21.37195$ for even-even nuclei; $c=-20.11223$ odd-A nuclei[76].

III. RESULTS AND DISCUSSION

The main purpose of this paper is to calculate the bifold potential (Coulomb potential) in a new type of nuclear density distribution. Based on the CPPM, the half-lives of α decay and cluster radioactivity of superheavy nuclei under various versions of proximity potentials are calculated. Within the framework of the CPPM, we calculated the half-lives of α decay and cluster radioactivity of 41 nuclei (211Fr to 242Cm) with 22 proximity potential forms. In addition, the results from Ni′s empirical formula and experimental data are listed in detail in Table 4. The first to third columns are the decay process, the energy released by the emitted nuclear decay, and the angular momentum taken away by the emitted nuclear, respectively. The last nine columns show the experimental data for cluster radioactivity and α decay half-lives, along with the results obtained from the CPPM using 22 different proximity potential forms and Nis empirical formula.It can be seen from this table that the results of the Prox.77 family, Prox.81 family, Bass family, Winther family and Ngô80 are within one order of magnitude of each other. Compared with the experimental data, the differences are less than 1 (except for a few nuclei), indicating that the theoretical results agree with the experimental datah. For Guo2013, the results generally differ from the experimental data by one to two orders of magnitude.

Table 4.Comparison of the discrepancy between the experimental cluster radioactivity and α decay half-lives and calculated ones using CPPM with 22 different versions of the proximity potential formalisms, the UDL, Ni's empirical formula, and the SL in alogarithmic form. The experimental cluster radioactivity and α decay half-lives are taken from Refs[42,45].

| *decay* | $Q_c$/MeV | $l$ | $\log_{10}^{T_{1/2}}/S$ | | | | | | | |
|---|---|---|---|---|---|---|---|---|---|---|
| | | | EXP | Prox.77-1 | Prox.77-2 | Prox.77-3 | Prox.77-4 | Prox.77-5 | Prox.77-6 | Prox.77-7 | Prox.77-8 |
| $^{221}$Fr→ $^{207}$Tl+$^{14}$C | 31.29 | 3 | 14.56 | 14.63 | 14.10 | 14.72 | 14.48 | 15.04 | 14.53 | 14.98 | 14.76 |
| $^{221}$Ra→ $^{207}$Pb+$^{14}$C | 32.39 | 3 | 13.39 | 13.22 | 12.68 | 13.30 | 13.06 | 13.64 | 13.12 | 13.57 | 13.35 |
| $^{222}$Ra→ $^{208}$Pb+$^{14}$C | 33.05 | 0 | 11.22 | 10.67 | 10.17 | 10.89 | 11.10 | 10.46 | 11.31 | 11.04 | 10.86 |
| $^{223}$Ra→ $^{209}$Pb+$^{14}$C | 31.83 | 4 | 15.06 | 14.39 | 13.86 | 14.48 | 14.22 | 14.80 | 14.29 | 14.74 | 14.51 |
| $^{224}$Ra→ $^{210}$Pb+$^{14}$C | 30.54 | 0 | 15.86 | 16.24 | 15.74 | 16.43 | 16.63 | 16.02 | 16.84 | 16.58 | 16.40 |
| $^{226}$Ra→ $^{212}$Pb+$^{14}$C | 28.19 | 0 | 21.19 | 21.96 | 21.46 | 22.12 | 22.32 | 21.73 | 22.51 | 22.27 | 22.10 |
| $^{223}$Ac→ $^{209}$Bi+$^{14}$C | 33.06 | 2 | 12.60 | 12.62 | 12.07 | 12.70 | 12.45 | 13.04 | 12.51 | 12.97 | 12.74 |
| $^{225}$Ac→ $^{211}$Bi+$^{14}$C | 30.48 | 4 | 17.16 | 18.45 | 17.92 | 18.53 | 18.63 | 18.85 | 18.35 | 18.79 | 18.56 |
| $^{226}$Th→ $^{212}$Po+$^{14}$C | 30.55 | 0 | >16.76 | 18.24 | 17.74 | 18.43 | 18.39 | 18.02 | 18.84 | 18.58 | 18.41 |
| $^{219}$Rn→ $^{215}$Po+$^{4}$He | 6.95 | 2 | 0.60 | 0.70 | 0.58 | 0.72 | 0.66 | 0.79 | 0.67 | 0.78 | 0.72 |
| $^{220}Rn \rightarrow ^{216}Po + ^{4}He$ | 6.40 | 0 | 1.75 | 2.43 | 2.32 | 2.48 | 2.53 | 2.38 | 2.57 | 2.49 | 2.47 |
| $^{221}Fr \rightarrow ^{217}At + ^{4}He$ | 6.46 | 2 | 2.46 | 2.84 | 2.72 | 2.87 | 2.81 | 2.94 | 2.82 | 2.93 | 2.87 |
| $^{223}Ra \rightarrow ^{219}Rn + ^{4}He$ | 5.98 | 2 | 5.99 | 5.16 | 5.05 | 5.18 | 5.12 | 5.52 | 5.14 | 5.24 | 5.19 |
| $^{226}Ra \rightarrow ^{222}Rn + ^{4}He$ | 4.87 | 0 | 10.70 | 10.28 | 10.15 | 10.30 | 10.34 | 10.21 | 10.38 | 10.33 | 10.29 |
| $^{223}Ac \rightarrow ^{219}Fr + ^{4}He$ | 6.78 | 2 | 2.10 | 2.47 | 2.34 | 2.49 | 2.43 | 2.56 | 2.45 | 2.55 | 2.49 |
| $^{227}Ac \rightarrow ^{223}Fr + ^{4}He$ | 5.04 | 0 | 10.70 | 10.02 | 9.91 | 10.04 | 9.99 | 10.11 | 10.00 | 10.09 | 10.05 |
| $^{226}Th \rightarrow ^{222}Po + ^{4}He$ | 6.45 | 0 | 3.27 | 3.85 | 3.74 | 3.90 | 3.95 | 3.81 | 4.00 | 3.94 | 3.89 |
| $^{227}Th \rightarrow ^{223}Ra + ^{4}He$ | 6.15 | 2 | 6.21 | 5.27 | 5.15 | 5.29 | 5.23 | 5.36 | 5.25 | 5.35 | 5.30 |
| $^{228}Th \rightarrow ^{224}Ra + ^{4}He$ | 5.52 | 0 | 7.78 | 7.86 | 7.74 | 7.89 | 7.94 | 7.80 | 7.98 | 7.93 | 7.89 |
| $^{229}Th \rightarrow ^{225}Po + ^{4}He$ | 5.17 | 2 | 11.40 | 9.76 | 9.65 | 9.78 | 9.73 | 9.85 | 9.74 | 9.83 | 9.78 |
| $^{232}Th \rightarrow ^{228}Ra + ^{4}He$ | 4.08 | 0 | 17.65 | 16.16 | 16.04 | 16.18 | 16.21 | 16.09 | 16.25 | 16.20 | 16.17 |
| $^{227}Pa \rightarrow ^{223}Ac + ^{4}He$ | 6.58 | 0 | 3.43 | 3.97 | 3.85 | 4.00 | 3.94 | 4.08 | 3.96 | 4.06 | 4.00 |
| $^{229}Pa \rightarrow ^{225}Ac + ^{4}He$ | 5.84 | 1 | 7.43 | 7.02 | 6.90 | 7.04 | 6.98 | 7.11 | 7.00 | 7.10 | 7.05 |
| $^{230}U \rightarrow ^{226}Th + ^{4}He$ | 5.99 | 0 | 6.24 | 6.58 | 6.46 | 6.62 | 6.67 | 6.53 | 6.71 | 6.66 | 6.61 |
| $^{232}U \rightarrow ^{228}Th + ^{4}He$ | 5.41 | 0 | 9.34 | 9.27 | 9.15 | 9.31 | 9.35 | 9.21 | 9.39 | 9.34 | 9.30 |
| $^{233}U \rightarrow ^{229}Th + ^{4}He$ | 4.91 | 0 | 12.70 | 12.23 | 12.12 | 12.25 | 12.20 | 12.31 | 12.21 | 12.30 | 12.26 |
| $^{236}U \rightarrow ^{232}Th + ^{4}He$ | 4.57 | 0 | 14.87 | 13.91 | 13.78 | 13.93 | 13.97 | 13.84 | 14.00 | 13.96 | 13.92 |
| $^{238}U \rightarrow ^{234}Th + ^{4}He$ | 4.27 | 0 | 17.15 | 15.84 | 15.71 | 15.85 | 15.89 | 15.77 | 15.93 | 15.88 | 15.84 |
| $^{231}Np \rightarrow ^{227}Pa + ^{4}He$ | 6.37 | 1 | 5.14 | 5.60 | 5.48 | 5.62 | 5.57 | 5.70 | 5.58 | 5.68 | 5.63 |
| $^{233}Np \rightarrow ^{229}Pa + ^{4}He$ | 5.63 | 0 | 8.49 | 8.85 | 8.73 | 8.87 | 8.81 | 8.93 | 8.83 | 8.92 | 8.87 |
| $^{235}Np \rightarrow ^{231}Pa + ^{4}He$ | 5.19 | 1 | 12.12 | 11.03 | 10.91 | 11.04 | 10.99 | 11.11 | 11.00 | 11.10 | 11.05 |
| $^{237}Np \rightarrow ^{233}Pa + ^{4}He$ | 4.96 | 1 | 13.83 | 12.25 | 12.15 | 12.27 | 12.22 | 12.33 | 12.23 | 12.34 | 12.28 |
| $^{237}Pu \rightarrow ^{233}U + ^{4}He$ | 5.75 | 1 | 6.60 | 8.66 | 8.55 | 8.68 | 8.63 | 8.76 | 8.65 | 8.74 | 8.69 |
| $^{239}Pu \rightarrow ^{235}U + ^{4}He$ | 5.24 | 0 | 11.88 | 11.19 | 11.17 | 11.21 | 11.16 | 11.28 | 11.17 | 11.27 | 11.22 |
| $^{237}Am \rightarrow ^{233}Np + ^{4}He$ | 6.20 | 1 | 7.24 | 7.09 | 6.96 | 7.11 | 7.05 | 7.18 | 7.07 | 7.17 | 7.11 |
| $^{239}Am \rightarrow ^{235}Np + ^{4}He$ | 5.92 | 1 | 8.63 | 8.30 | 8.18 | 8.32 | 8.27 | 8.40 | 8.28 | 8.38 | 8.33 |
| $^{241}Am \rightarrow ^{237}Np + ^{4}He$ | 5.64 | 1 | 10.14 | 9.60 | 9.47 | 9.61 | 9.56 | 9.68 | 9.57 | 9.67 | 9.62 |
| $^{240}Cm \rightarrow ^{236}Pu + ^{4}He$ | 6.40 | 0 | 6.42 | 6.41 | 6.29 | 6.46 | 6.51 | 6.36 | 6.55 | 6.50 | 6.45 |
| $^{241}Cm \rightarrow ^{237}Np + ^{4}He$ | 6.19 | 3 | 8.45 | 7.49 | 7.37 | 7.51 | 7.46 | 7.59 | 7.47 | 7.57 | 7.52 |
| $^{243}Cm \rightarrow ^{239}Np + ^{4}He$ | 6.17 | 2 | 8.96 | 7.56 | 7.44 | 7.58 | 7.53 | 7.66 | 7.54 | 7.64 | 7.59 |
| $^{244}Cm \rightarrow ^{240}Np + ^{4}He$ | 5.90 | 0 | 8.76 | 8.56 | 8.44 | 8.61 | 8.65 | 8.51 | 8.70 | 8.64 | 8.60 |

Table 4-continued from previous page

| *decay* | $Q_c$/MeV | $l$ | $\log_{10}^{T_{1/2}}/S$ | | | | | | | | |
|---|---|---|---|---|---|---|---|---|---|---|---|
| | | | EXP | Prox.77-1 | Prox.77-2 | Prox.77-3 | Prox.77-4 | Prox.77-5 | Prox.77-6 | Prox.77-7 | Prox.77-8 |
| $^{221}Fr \rightarrow ^{207}Tl + ^{14}C$ | 31.29 | 3 | 14.56 | 14.89 | 14.40 | 14.72 | 14.48 | 14.31 | 14.67 | 15.47 | 14.82 |

| | | | | | | | | | | |
|---|---|---|---|---|---|---|---|---|---|---|
| $^{221}Ra \to ^{207}Pb + ^{14}C$ | 32.39 | 3 | 13.39 | 13.48 | 12.99 | 13.30 | 13.06 | 12.89 | 13.26 | 14.08 | 13.40 |
| $^{222}Ra \to ^{208}Pb + ^{14}C$ | 33.05 | 0 | 11.22 | 11.09 | 10.58 | 10.89 | 11.10 | 10.52 | 11.00 | 11.34 | 11.05 |
| $^{223}Ra \to ^{209}Pb + ^{14}C$ | 31.83 | 4 | 15.06 | 14.65 | 14.16 | 14.48 | 14.22 | 14.06 | 14.42 | 15.24 | 14.57 |
| $^{224}Ra \to ^{210}Pb + ^{14}C$ | 30.54 | 0 | 15.86 | 16.36 | 16.14 | 16.43 | 16.63 | 16.08 | 16.54 | 16.88 | 16.59 |
| $^{226}Ra \to ^{212}Pb + ^{14}C$ | 28.19 | 0 | 21.19 | 22.31 | 21.85 | 22.12 | 22.32 | 21.78 | 22.22 | 22,56 | 22.28 |
| $^{223}Ac \to ^{209}Bi + ^{14}C$ | 33.06 | 2 | 12.60 | 12.88 | 12.37 | 15.07 | 14.26 | 12.28 | 12.65 | 13.48 | 12.80 |
| $^{225}Ac \to ^{211}Bi + ^{14}C$ | 30.48 | 4 | 17.16 | 18.70 | 18.22 | 13.66 | 12.84 | 18.13 | 18.48 | 19.28 | 18.62 |
| $^{226}Th \to ^{212}Po + ^{14}C$ | 30.55 | 0 | >16.76 | 18.63 | 18.14 | 10.44 | 10.36 | 18.08 | 18.54 | 18.88 | 18.59 |
| $^{219}Rn \to ^{215}Po + ^{4}He$ | 6.95 | 2 | 0.60 | 0.76 | 0.64 | 14.83 | 14.01 | 0.63 | 0.70 | 0.93 | 1.33 |
| $^{220}Rn \to ^{216}Po + ^{4}He$ | 6.40 | 0 | 1.75 | 2.52 | 2.41 | 16.01 | 15.92 | 2.44 | 2.50 | 2.62 | 3.09 |
| $^{221}Fr \to ^{217}At + ^{4}He$ | 6.46 | 2 | 2.46 | 2.90 | 2.79 | 21.72 | 21.63 | 2.40 | 2.85 | 3.08 | 3.45 |
| $^{223}Ra \to ^{219}Rn + ^{4}He$ | 5.98 | 2 | 5.99 | 5.22 | 5.11 | 13.06 | 12.23 | 5.09 | 5.17 | 5.39 | 5.75 |
| $^{226}Ra \to ^{222}Rn + ^{4}He$ | 4.87 | 0 | 10.70 | 10.34 | 10.24 | 18.87 | 18.08 | 10.23 | 10.32 | 10.44 | 10.57 |
| $^{223}Ac \to ^{219}Fr + ^{4}He$ | 6.78 | 2 | 2.10 | 2.52 | 2.41 | 18.01 | 17.92 | 2.39 | 2.47 | 1.70 | 3.09 |
| $^{227}Ac \to ^{223}Fr + ^{4}He$ | 5.04 | 0 | 10.70 | 10.08 | 9.98 | 0.80 | 0.61 | 9.96 | 10.03 | 10.24 | 10.57 |
| $^{226}Th \to ^{222}Po + ^{4}He$ | 6.45 | 0 | 3.27 | 3.95 | 3.83 | 3.81 | 3.78 | 3.82 | 3.93 | 4.05 | 4.51 |
| $^{227}Th \to ^{223}Ra + ^{4}He$ | 6.15 | 2 | 6.21 | 5.33 | 5.22 | 5.37 | 5.19 | 5.20 | 5.28 | 5.50 | 5.86 |
| $^{228}Th \to ^{224}Ra + ^{4}He$ | 5.52 | 0 | 7.78 | 7.94 | 7.83 | 7.80 | 7.78 | 7.82 | 7.92 | 8.03 | 8.46 |
| $^{229}Th \to ^{225}Po + ^{4}He$ | 5.17 | 2 | 11.40 | 9.81 | 9.71 | 9.85 | 9.68 | 9.69 | 9.77 | 9.98 | 10.30 |
| $^{232}Th \to ^{228}Ra + ^{4}He$ | 4.08 | 0 | 17.65 | 16.21 | 16.12 | 16.10 | 16.08 | 16.11 | 16.19 | 16.30 | 16.66 |
| $^{227}Pa \to ^{223}Ac + ^{4}He$ | 6.58 | 0 | 3.43 | 4.03 | 3.92 | 4.07 | 3.89 | 3.90 | 3.99 | 4.21 | 4.59 |
| $^{229}Pa \to ^{225}Ac + ^{4}He$ | 5.84 | 1 | 7.43 | 7.08 | 6.97 | 7.12 | 6.94 | 6.95 | 7.03 | 7.25 | 7.60 |
| $^{230}U \to ^{226}Th + ^{4}He$ | 5.99 | 0 | 6.24 | 6.67 | 6.55 | 6.52 | 6.50 | 6.54 | 6.65 | 6.76 | 7.21 |
| $^{232}U \to ^{228}Th + ^{4}He$ | 5.41 | 0 | 9.34 | 9.35 | 9.24 | 9.21 | 9.19 | 9.23 | 9.33 | 9.44 | 9.86 |
| $^{233}U \to ^{229}Th + ^{4}He$ | 4.91 | 0 | 12.70 | 12.29 | 12.19 | 12.31 | 12.16 | 12.17 | 12.23 | 12.45 | 12.76 |
| $^{236}U \to ^{232}Th + ^{4}He$ | 4.57 | 0 | 14.87 | 13.96 | 13.86 | 13.84 | 13.82 | 13.85 | 13.94 | 14.06 | 14.44 |
| $^{238}U \to ^{234}Th + ^{4}He$ | 4.27 | 0 | 17.15 | 15.88 | 15.79 | 15.77 | 15.75 | 15.78 | 15.87 | 15.98 | 16.34 |
| $^{231}Np \to ^{227}Pa + ^{4}He$ | 6.37 | 1 | 5.14 | 5.66 | 5.55 | 5.70 | 5.52 | 5.52 | 5.61 | 5.83 | 6.21 |
| $^{233}Np \to ^{229}Pa + ^{4}He$ | 5.63 | 0 | 8.49 | 8.90 | 8.79 | 8.94 | 8.76 | 8.77 | 8.85 | 9.07 | 9.41 |
| $^{235}Np \to ^{231}Pa + ^{4}He$ | 5.19 | 1 | 12.12 | 11.08 | 10.98 | 11.11 | 10.95 | 10.96 | 11.03 | 11.25 | 11.57 |
| $^{237}Np \to ^{233}Pa + ^{4}He$ | 4.96 | 1 | 13.83 | 12.31 | 12.21 | 12.34 | 12.18 | 12.19 | 12.26 | 12.47 | 12.79 |
| $^{237}Pu \to ^{233}U + ^{4}He$ | 5.75 | 1 | 6.60 | 8.72 | 8.62 | 8.76 | 8.59 | 8.60 | 8.67 | 8.89 | 9.24 |
| $^{239}Pu \to ^{235}U + ^{4}He$ | 5.24 | 0 | 11.88 | 11.25 | 11.15 | 11.28 | 11.11 | 11.12 | 11.20 | 11.41 | 11.74 |
| $^{237}Am \to ^{233}Np + ^{4}He$ | 6.20 | 1 | 7.24 | 7.15 | 7.04 | 7.19 | 7.01 | 7.02 | 7.10 | 7.32 | 7.68 |
| $^{239}Am \to ^{235}Np + ^{4}He$ | 5.92 | 1 | 8.63 | 8.36 | 8.25 | 8.39 | 8.22 | 8.23 | 8.31 | 8.53 | 8.88 |
| $^{241}Am \to ^{237}Np + ^{4}He$ | 5.64 | 1 | 10.14 | 9.65 | 9.64 | 9.69 | 9.51 | 9.52 | 9.60 | 9.82 | 10.16 |

| | | | | | | | | | | |
|---|---|---|---|---|---|---|---|---|---|---|
| $^{240}Cm\to^{236}Pu+^{4}He$ | 6.40 | 0 | 6.42 | 6.51 | 6.39 | 6.36 | 6.34 | 6.38 | 6.49 | 6.60 | 7.06 |
| $^{241}Cm\to^{237}Np+^{4}He$ | 6.19 | 3 | 8.45 | 7.55 | 7.44 | 7.59 | 7.41 | 7.42 | 7.50 | 7.72 | 8.07 |
| $^{243}Cm\to^{239}Np+^{4}He$ | 6.17 | 2 | 8.96 | 7.62 | 7.51 | 7.66 | 7.47 | 7.49 | 7.57 | 7.79 | 8.15 |
| $^{244}Cm\to^{240}Np+^{4}He$ | 5.90 | 0 | 8.76 | 8.64 | 8.50 | 8.51 | 8.48 | 8.52 | 8.63 | 8.74 | 9.18 |

Table 4-continued from previous page

| *decay* | $Q_c$/MeV | $l$ | $\log_{10}^{T_{1/2}}/S$ | | | | | | | | |
|---|---|---|---|---|---|---|---|---|---|---|---|
| | | | EXP | Bass80 | CW76 | BW91 | AW95 | Ngô80 | Guo2013 | Ni | |
| $^{221}Fr\to^{207}Tl+^{14}C$ | 31.29 | 3 | 14.56 | 15.15 | 14.55 | 14.32 | 14.37 | 14.64 | 16.06 | 14.63 | |
| $^{221}Ra\to^{207}Pb+^{14}C$ | 32.39 | 3 | 13.39 | 13.74 | 13.14 | 12.90 | 13.46 | 13.34 | 14.67 | 13.48 | |
| $^{222}Ra\to^{208}Pb+^{14}C$ | 33.05 | 0 | 11.22 | 11.28 | 10.79 | 10.68 | 11.55 | 11.16 | 12.28 | 11.02 | |
| $^{223}Ra\to^{209}Pb+^{14}C$ | 31.83 | 4 | 15.06 | 14.91 | 14.31 | 14.06 | 14.63 | 14.46 | 15.83 | 14.56 | |
| $^{224}Ra\to^{210}Pb+^{14}C$ | 30.54 | 0 | 15.86 | 16.80 | 16.34 | 16.23 | 17.06 | 16.69 | 16.80 | 15.86 | |
| $^{226}Ra\to^{212}Pb+^{14}C$ | 28.19 | 0 | 21.19 | 22.48 | 22.03 | 21.94 | 22.72 | 22.37 | 22.48 | 20.94 | |
| $^{223}Ac\to^{209}Bi+^{14}C$ | 33.06 | 2 | 12.60 | 13.14 | 12.53 | 12.28 | 12.85 | 12.82 | 14.09 | 13.19 | |
| $^{225}Ac\to^{211}Bi+^{14}C$ | 30.48 | 4 | 17.16 | 18.94 | 18.36 | 18.13 | 18.66 | 18.40 | 19.86 | 18.24 | |
| $^{226}Th\to^{212}Po+^{14}C$ | 30.55 | 0 | >16.76 | 18.80 | 18.34 | 18.23 | 19.05 | 18.71 | 17.54 | 17.82 | |
| $^{219}Rn\to^{215}Po+^{4}He$ | 6.95 | 2 | 0.60 | 1.08 | 1.12 | 1.07 | 1.07 | 2.23 | 1.00 | 0.55 | |
| $^{220}Rn\to^{216}Po+^{4}He$ | 6.40 | 0 | 1.75 | 2.82 | 2.89 | 2.87 | 2.93 | 2.61 | 2.54 | 1.93 | |
| $^{221}Fr\to^{217}At+^{4}He$ | 6.46 | 2 | 2.46 | 3.21 | 3.25 | 3.21 | 3.20 | 4.32 | 3.13 | 2.71 | |
| $^{223}Ra\to^{219}Rn+^{4}He$ | 5.98 | 2 | 5.99 | 5.52 | 5.56 | 5.51 | 5.50 | 6.58 | 5.44 | 5.39 | |
| $^{226}Ra\to^{222}Rn+^{4}He$ | 4.87 | 0 | 10.70 | 10.59 | 10.67 | 10.65 | 10.69 | 10.42 | 10.35 | 10.67 | |
| $^{223}Ac\to^{219}Fr+^{4}He$ | 6.78 | 2 | 2.10 | 2.84 | 2.88 | 2.84 | 2.83 | 3.99 | 2.76 | 2.27 | |
| $^{227}Ac\to^{223}Fr+^{4}He$ | 5.04 | 0 | 10.70 | 10.35 | 10.39 | 10.34 | 10.34 | 11.36 | 10.27 | 10.73 | |
| $^{226}Th\to^{222}Po+^{4}He$ | 6.45 | 0 | 3.27 | 4.25 | 4.31 | 4.30 | 4.35 | 4.04 | 3.96 | 3.44 | |
| $^{227}Th\to^{223}Ra+^{4}He$ | 6.15 | 2 | 6.21 | 5.63 | 5.67 | 5.63 | 5.61 | 6.72 | 5.55 | 5.51 | |
| $^{228}Th\to^{224}Ra+^{4}He$ | 5.52 | 0 | 7.78 | 8.21 | 8.27 | 8.26 | 8.30 | 8.02 | 7.94 | 7.88 | |
| $^{229}Th\to^{225}Po+^{4}He$ | 5.17 | 2 | 11.40 | 10.09 | 10.12 | 10.08 | 10.07 | 11.09 | 10.01 | 10.61 | |
| $^{232}Th\to^{228}Ra+^{4}He$ | 4.08 | 0 | 17.65 | 16.44 | 16.50 | 16.49 | 16.52 | 16.28 | 16.21 | 17.57 | |
| $^{227}Pa\to^{223}Ac+^{4}He$ | 6.58 | 0 | 3.43 | 4.35 | 4.39 | 4.35 | 4.33 | 5.51 | 4.27 | 3.92 | |
| $^{229}Pa\to^{225}Ac+^{4}He$ | 5.84 | 1 | 7.43 | 7.37 | 7.41 | 7.37 | 7.45 | 8.46 | 7.30 | 7.30 | |
| $^{230}U\to^{226}Th+^{4}He$ | 5.99 | 0 | 6.24 | 6.95 | 7.01 | 7.00 | 7.05 | 6.75 | 6.67 | 6.41 | |
| $^{232}U\to^{228}Th+^{4}He$ | 5.41 | 0 | 9.34 | 9.61 | 9.68 | 9.66 | 9.71 | 9.43 | 9.35 | 9.46 | |
| $^{233}U\to^{229}Th+^{4}He$ | 4.91 | 0 | 12.70 | 12.56 | 12.59 | 12.56 | 12.54 | 13.55 | 12.48 | 13.25 | |

| | | | | | | | | | | |
|---|---|---|---|---|---|---|---|---|---|---|
| $^{236}U \rightarrow ^{232}Th + ^{4}He$ | 4.57 | 0 | 14.87 | 14.20 | 14.27 | 14.25 | 14.29 | 14.04 | 13.97 | 14.86 | |
| $^{238}U \rightarrow ^{234}Th + ^{4}He$ | 4.27 | 0 | 17.15 | 16.12 | 16.18 | 16.17 | 16.20 | 15.96 | 15.89 | 17.17 | |
| $^{231}Np \rightarrow ^{227}Pa + ^{4}He$ | 6.37 | 1 | 5.14 | 5.96 | 6.01 | 5.96 | 5.95 | 7.11 | 5.89 | 6.69 | |
| $^{233}Np \rightarrow ^{229}Pa + ^{4}He$ | 5.63 | 0 | 8.49 | 9.18 | 9.23 | 9.19 | 9.17 | 10.27 | 9.12 | 9.33 | |
| $^{235}Np \rightarrow ^{231}Pa + ^{4}He$ | 5.19 | 1 | 12.12 | 11.35 | 11.40 | 11.35 | 11.33 | 12.39 | 11.28 | 11.85 | |
| $^{237}Np \rightarrow ^{233}Pa + ^{4}He$ | 4.96 | 1 | 13.83 | 12.57 | 12.62 | 12.58 | 12.55 | 13.59 | 12.51 | 13.31 | |
| $^{237}Pu \rightarrow ^{233}U + ^{4}He$ | 5.75 | 1 | 6.60 | 9.01 | 9.05 | 9.01 | 8.99 | 10.11 | 8.94 | 9.31 | |
| $^{239}Pu \rightarrow ^{235}U + ^{4}He$ | 5.24 | 0 | 11.88 | 11.52 | 11.57 | 11.52 | 11.49 | 12.58 | 11.46 | 12.20 | |
| $^{237}Am \rightarrow ^{233}Np + ^{4}He$ | 6.20 | 1 | 7.24 | 7.44 | 7.50 | 7.45 | 7.42 | 8.59 | 7.38 | 7.36 | |
| $^{239}Am \rightarrow ^{235}Np + ^{4}He$ | 5.92 | 1 | 8.63 | 8.64 | 8.69 | 8.65 | 8.62 | 9.77 | 8.59 | 8.74 | |
| $^{241}Am \rightarrow ^{237}Np + ^{4}He$ | 5.64 | 1 | 10.14 | 9.93 | 9.98 | 9.94 | 9.91 | 11.02 | 9.87 | 10.23 | |
| $^{240}Cm \rightarrow ^{236}Pu + ^{4}He$ | 6.40 | 0 | 6.42 | 6.78 | 6.87 | 6.85 | 6.89 | 6.61 | 6.53 | 6.26 | |
| $^{241}Cm \rightarrow ^{237}Np + ^{4}He$ | 6.19 | 3 | 8.45 | 7.84 | 7.89 | 7.85 | 7.82 | 8.95 | 7.78 | 8.00 | |
| $^{243}Cm \rightarrow ^{239}Np + ^{4}He$ | 6.17 | 2 | 8.96 | 7.91 | 7.97 | 7.92 | 7.89 | 9.05 | 7.86 | 8.10 | |
| $^{244}Cm \rightarrow ^{240}Np + ^{4}He$ | 5.90 | 0 | 8.76 | 8.91 | 9.00 | 8.98 | 9.01 | 8.47 | 8.67 | 8.71 | |

In order to compare results of radioactive half-lives of the cluster(or α decay) with the experimental data, we calculated the RMS deviation σ which is defined as:

$$\sigma = \sqrt{\sum_{i=1}^{n} \frac{(\log_{10}^{T_{1/2}^{\exp,i}} - \log_{10}^{T_{1/2}^{cal,i}})^2}{n}}, \quad (24)$$

where, $\log_{10}^{T_{1/2}^{\exp,i}}$ denotes the logarithmic form of theexperimental value of the cluster radioactivity and α decay half-lives of the i-th nucleus. $\log_{10}^{T_{1/2}^{cal,i}}$ denotes the logarithmic form of the calculated value of the cluster radioactive and α decay half-lives of the i-th nucleus corresponding to the experimental data. This research involves the decays of a total of 41 atomic nuclei. Table 5 lists the detailed results of the RMS deviation of 22 different versions of proximity potential formalisms and Ni's empirical formula. It can be found that BW91 in the Winther family has the lowest RMS deviation ($\sigma = 0.812$), therefore BW91 is the most suitable form of proximity potential. Simultaneously, the RMS of CW76 ($\sigma = 0.816$) in the Winther family is very close to the RMS of BW91. Among the 22 different versions of proximity potentials, except version Ng ô80, the RMS deviations obtained from other versions of proximity potentials are less than 1. These versions can be applied to cluster radioactivity and α decay of superheavy nuclei. In addition the results of the proximity potential calculation data in the form of Guo2013 show a large deviation from the experimental value. In general, the half-lives of nuclear decay calculated using the CPPM exhibit a high degree of accuracy when compared to experimental half-life values.

Table 5. RMS deviation σ between the experimental data and calculated ones using CPPM with 22 different versions of the proximity potential formalisms, and Ni's empirical formula for cluster radioactivity.

| Method | Prox.77-1 | Prox.77-2 | Prox.77-3 | Prox.77-4 | Prox.77-5 | Prox.77-6 | Prox.77-7 | Prox.77-8 |
|---|---|---|---|---|---|---|---|---|
| σ | 0.843 | 0.881 | 0.846 | 0.85 | 0.841 | 0.898 | 0.867 | 0.851 |
| Method | Prox.77-9 | Prox.77-10 | Prox.77-11 | Prox.77-12 | Prox.77-13 | Prox.81 | Bass73 | Bass77 |
| σ | 0.864 | 0.851 | 0.841 | 0.861 | 0.858 | 0.867 | 0.924 | 0.88 |
| Method | Bass80 | CW76 | BW91 | AW95 | Ngô80 | Guo2013 | Ni | |
| σ | 0.884 | 0.816 | 0.812 | 0.905 | 1.155 | 0.982 | 0.646 | |

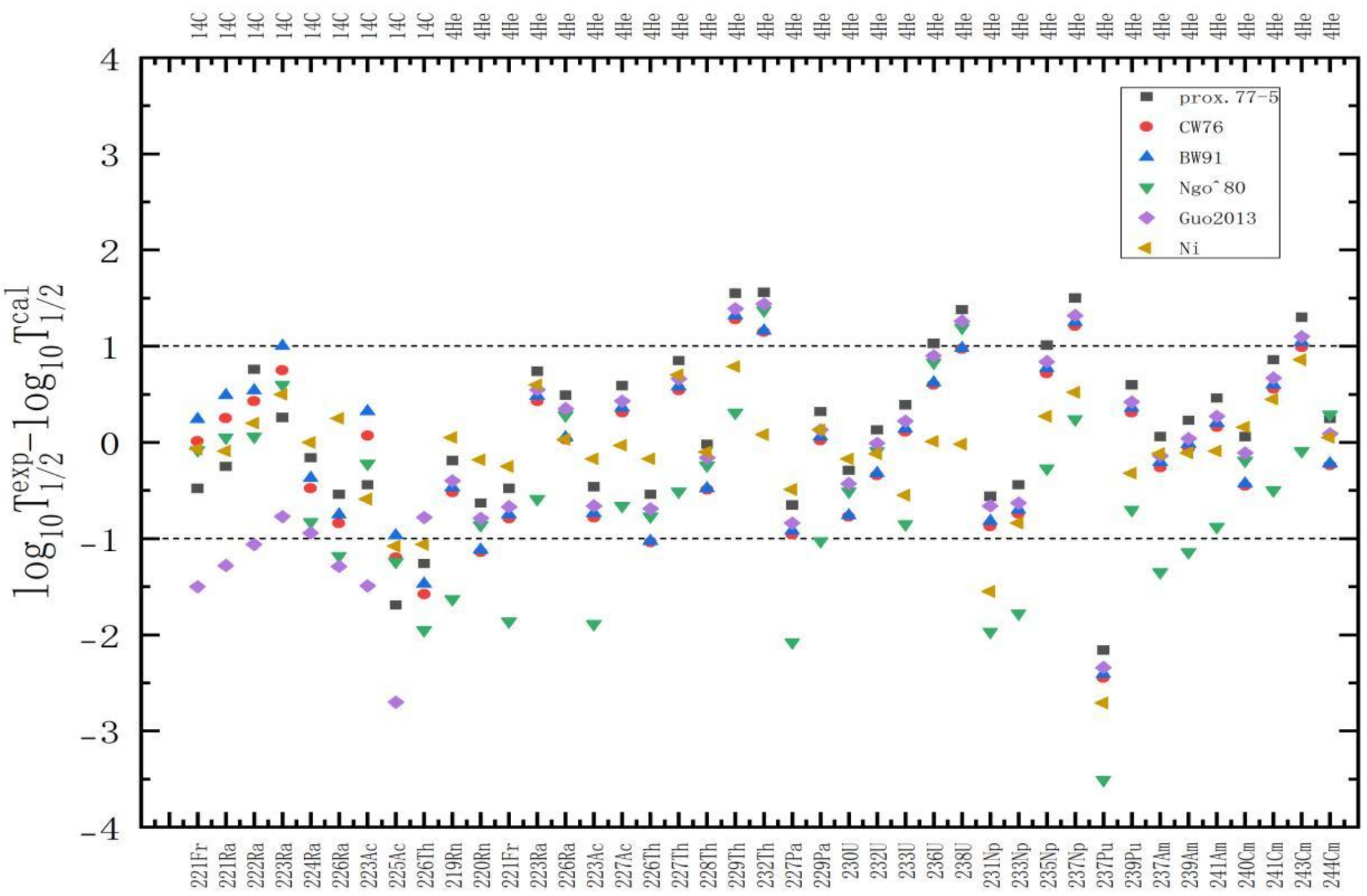

FIG.1.Comparison of experimental and calculated half-lives for cluster radioactivity and α decay in the CPPM framework using Prox.77-5, CW76, BW91, Ngoˆ80 and Guo2013, as proximity potentials, shown in logarithmic form.

To further verify the feasibility of applying proximity potential to cluster radioactivity and α decay, we compare experimental data of cluster radioactivity and α decay half-lives with the half-lives of CPPM with proximity potential Prox.77-5, CW76, BW91, Ngô80, Guo2013 and Ni's empirical formula, in FIG.1. Under the framework of the CPPM, the results of the half-lives of cluster radioactivity and α decay calculated using the proximity potentials of CW76, BW91 and Prox.77-5 are highly consistent. However, there are relatively large deviations in the results of the half-lives of cluster radioactivity and α decay calculated using the proximity potentials of Ngô80 and Guo2013 within the framework of the CPPM.

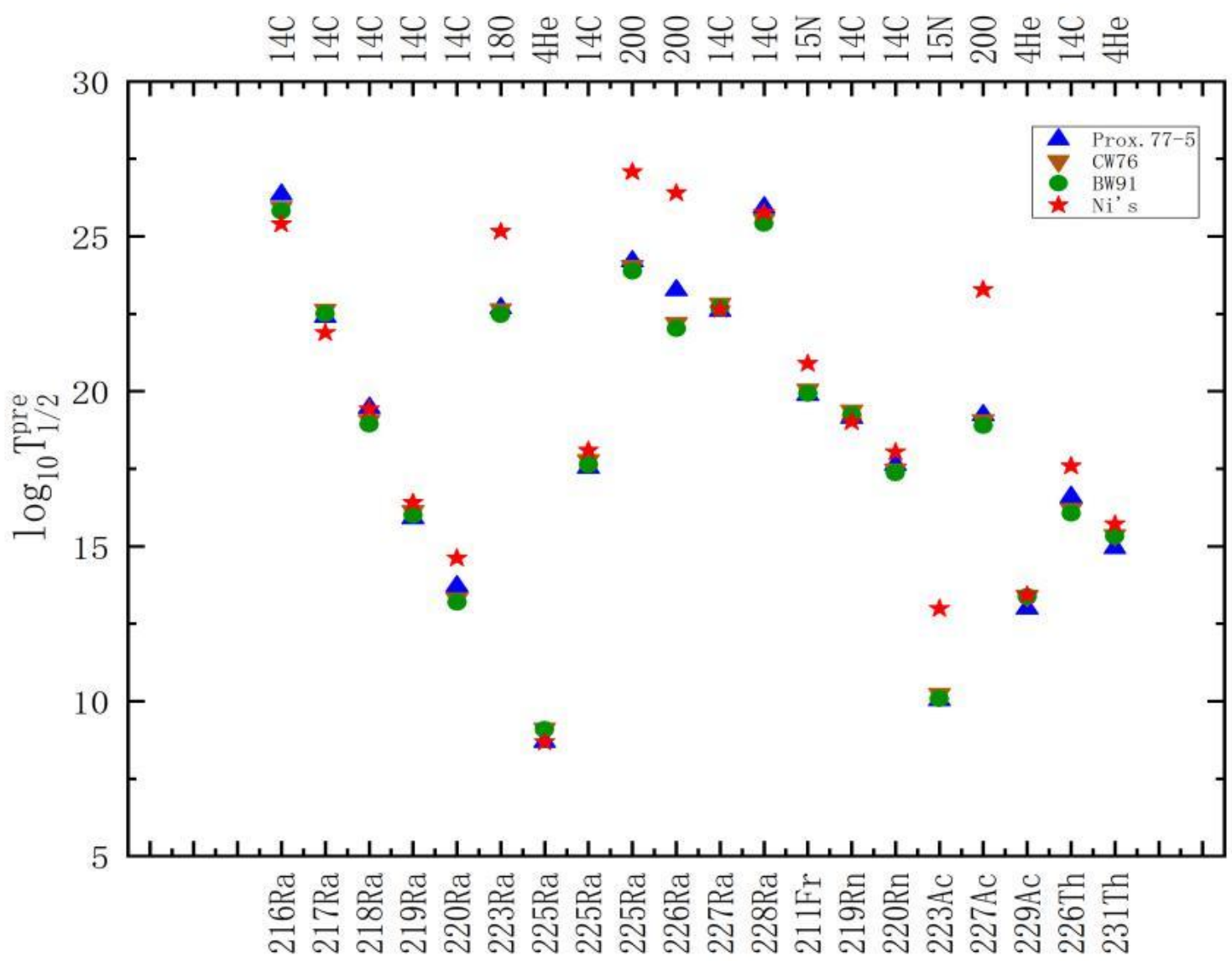

FIG. 2. The predicted half-lives were calculater using CPPM with Prox.77-5, CW76, BW91 and Ni's empirical formula, presented in logarithmic form.

Table 6.Predicted half-lives for 20 possible cluster radioactivity and α decay nuclei

| *decay* | $Q_c$/MeV | *l* | $\log_{10}^{T_{1/2}}/S$ | | | | |
|---|---|---|---|---|---|---|---|
| | | | EXP | Prox.77-5 | CW76 | BW91 | Ni's |
| $^{216}Ra \to ^{202}Pb + ^{14}C$ | 26.32 | 0 | \ | 26.35 | 25.93 | 25.82 | 25.40 |
| $^{217}Ra \to ^{203}Pb + ^{14}C$ | 27.76 | 3 | \ | 22.41 | 22.63 | 22.50 | 21.89 |
| $^{218}Ra \to ^{204}Pb + ^{14}C$ | 28.84 | 0 | \ | 19.47 | 19.06 | 18.94 | 19.43 |
| $^{219}Ra \to ^{205}Pb + ^{14}C$ | 30.25 | 1 | \ | 15.91 | 16.14 | 16.01 | 16.41 |
| $^{220}Ra \to ^{206}Pb + ^{14}C$ | 31.14 | 0 | \ | 13.71 | 13.32 | 13.19 | 14.62 |
| $^{223}Ra \to ^{205}Hg + ^{18}O$ | 40.30 | 1 | \ | 22.68 | 22.65 | 22.48 | 25.15 |
| $^{225}Ra \to ^{221}Rn + ^{4}He$ | 5.10 | 4 | \ | 8.69 | 9.11 | 9.08 | 8.68 |
| $^{225}Ra \to ^{211}Pb + ^{14}C$ | 29.47 | 4 | \ | 17.54 | 17.77 | 17.64 | 18.09 |
| $^{225}Ra \to ^{205}Hg + ^{20}O$ | 40.48 | 1 | \ | 24.20 | 24.04 | 23.87 | 27.08 |
| $^{226}Ra \to ^{206}Hg + ^{20}O$ | 40.82 | 0 | \ | 23.26 | 22.20 | 22.02 | 26.40 |
| $^{227}Ra \to ^{213}Pb + ^{14}C$ | 27.47 | 4 | \ | 22.60 | 22.83 | 22.70 | 22.64 |
| $^{228}Ra \to ^{214}Pb + ^{14}C$ | 26.21 | 0 | \ | 25.94 | 25.54 | 25.42 | 25.77 |
| $^{221}Fr \to ^{206}Hg + ^{15}N$ | 34.12 | 3 | \ | 19.89 | 20.06 | 19.93 | 20.89 |
| $^{219}Rn \to ^{205}Hg + ^{14}C$ | 28.10 | 3 | \ | 19.15 | 19.37 | 19.24 | 19.02 |
| $^{220}Rn \to ^{206}Hg + ^{14}C$ | 28.64 | 0 | \ | 17.63 | 17.48 | 17.36 | 18.04 |
| $^{223}Ac \to ^{208}Pb + ^{15}N$ | 39.60 | 3 | \ | 10.04 | 10.23 | 10.08 | 12.99 |
| $^{227}Ac \to ^{207}Tl + ^{20}O$ | 43.09 | 1 | \ | 19.23 | 19.07 | 18.89 | 23.28 |
| $^{229}Ac \to ^{225}Fr + ^{4}He$ | 4.44 | 1 | \ | 12.99 | 13.40 | 13.38 | 13.42 |
| $^{226}Th \to ^{212}Po + ^{14}C$ | 30.66 | 0 | \ | 16.59 | 16.20 | 16.07 | 17.59 |
| $^{231}Th \to ^{227}Ra + ^{4}He$ | 4.21 | 2 | \ | 14.95 | 15.36 | 15.33 | 15.71 |

Therefore, we have employed the four nearest-neighbor potentials with the smallest RMS deviations to predict the half-lives of 20 possible candidates for cluster radioactivity and α decay. In Table 6, the first to the third columns are the decay process, the energy released by the cluster radioactivity and α decay of the emitted nucleus and the angular momentum taken away by the emitted nucleus, respectively. The fourth column represents the experimental data of cluster radioactivity and α decay half-lives.The last four columns represent the predicted results in a logarithmic form of CPPM with Prox.77-5, CW76, BW91 and Ni's empirical formula, respectively. The decay energies and experimental half-lives of α decay and cluster radioactivity are taken from AME2020[56] and NUBASE2020[43]. In Table 6, the prediction results of our model (CPPM with Prox.77-5, CW76 and BW91) are basically consistent with the results from the Ni's empirical formula. This indicates that using our model to predict the half-lives of cluster radioactivity and α decay is reliable. And it will be helpful for searching for new cluster radioactivity and α decay in future experiments.

In FIG.2, we plotted the half-lives calculated by using three forms of proximity potential with the minimum RMS deviation and the empirical formula of Ni within the framework of the CPPM. This can intuitively reflect the consistency between our prediction results and the empirical formula of Ni in terms of cluster radioactivity and α decay. In FIG.2, the prediction results of cluster radioactivity and α decay half-lives obtained from the CPPM with three proximity potentials and the Ni empirical formula are very close. This indicates that the prediction results of this study have relatively high reliability.

## IV. SUMMARY

We have investigated 22 versions of the proximity potential for the nuclear potential part that can be used to study the half-lives of 41 cluster radioactivity and α decay processes. The RMS deviation method is used to compare the theoretical results with the experimental data. The research results show that CPPM with the proximity potential BW91 is suitable for studying cluster radioactivity and α decay half-lives. In the CPPM with the proximity potentials CW76 and Prox.77-5, the RMS deviation is small. In addition, we use the CPPM with four proximity potentials to predict the half-lives of 20 candidate clusters. These results are in good agreement with the results from Ni's semi-empirical formula. This provides a reliable tool for predicting cluster radioactivity and α decay half-lives in superheavy nuclei.

V. ACKNOWLEDGEMENTS

This work is supported by Yunnan Provincial Science Foundation Project (No. 202501AT070067), Yunnan Provincial Xing Dian Talent Support Program (Young Talents Special Program), Kunming University Talent Introduction Research Project (No. YJL24019), Yunnan Provincial Department of Education Scientific Research Fund Project (No. 2025Y1055 and 2025Y1042), the Program for Frontier Research Team of Kunming University 2023, and National Natural Science Foundation of China (No. 12063006).